\documentclass[conference]{IEEEtran}
\IEEEoverridecommandlockouts
\usepackage{cite}
\usepackage{amsmath,amssymb,amsfonts}
\usepackage{algorithmic}
\usepackage{graphicx}
\usepackage{textcomp}
\usepackage{xcolor}
\usepackage{enumitem} 
\def\BibTeX{{\rm B\kern-.05em{\sc i\kern-.025em b}\kern-.08em
    T\kern-.1667em\lower.7ex\hbox{E}\kern-.125emX}}
\begin{document}
\title{Blockchain and BIM (Building Information Modeling): Progress in Academia and Industry}

\author{
	\IEEEauthorblockN{Michael Kuperberg\\}
	\IEEEauthorblockA{
		\textit{Blockchain and Distributed Ledgers Group} \\
		\textit{DB Systel GmbH}\\ 
			Frankfurt am Main, Germany \\
			michael.kuperberg@deutschebahn.com
	}
\and
	\IEEEauthorblockN{Matthias Geipel\\}
	\IEEEauthorblockA{
		\textit{Advisory Services} \\
		\textit{Arup Deutschland GmbH}\\ 
			Berlin, Germany \\
			matthias.geipel@arup.com}
}

\maketitle

\begin{abstract}
In construction, BIM (Building Information Modeling) promises to increase quality of data and to provide a shared, uniform view to all parties. 
While BIM tools and exchange formats exist, the distribution and safeguarding of data is an ongoing challenge. 
Distributed Ledger Technology and Blockchains offer a possible solution to this task, and they promise quality attributes such as tamper resistance,  traceability/auditability and safe digitalization of assets and intellectual property.  
However, the practical application and adoption of Distributed Ledger Technology in the built environment requires a good understanding of tool maturity, performance and standardization. 
Also, user-oriented integration of BIM tools 
with the blockchain backend needs attention. 
The contribution of this paper is an overview over both industrial and academic progress at the intersection of BIM and blockchains/DLT. 

\end{abstract}


\section{Introduction and Problem Statement}
\label{Introduction}
Building Information Modeling (BIM) is an 
IT-based, 
data-driven 
and process-centric approach to design, construction, operation, maintenance and disposal of buildings. 
Construction processes naturally involve different stakeholders, with varying and often conflicting interests and motivations. 
Therefore, transparency, auditability, accountability and tamper-resistance are highly desired qualities in the BIM projects, and the increased market adoption of BIM means that a more fluent exchange of data (and thus data reconciliation) bear a significant economic potential. 
In large-scale projects, BIM is often mandated by law or by the client/customer (see e.g. \cite{BIM-bahn}). 

One technology that can enable BIM processes in assuring these quality aspects is Distributed Ledger Technology (DLT). 
A DLT is essentially a replicated, highly available, distributed database which enables different partners to maintain a trusted, shared view on data. 
Traditionally, ledgers are append-only data structures which, by design, keep a full history of changes. 
In a DLT implementation, the same logical ledger (``common truth'') is available in multiple physical copies, which are kept synchronized. 
Blockchain technology is a type of DLT 
\textcolor{black}
{that} writes append-only data into sequential blocks which are concatenated through hashes. 
Additionally, blockchain implementations employ consensus algorithms to ensure that all parties (or a sufficient majority thereof) agree to the data before it is added to the ledger.  
In particular, blockchains utilize decentralization: a blockchain network is not governed by a ``master'' party, and data cannot be overwritten/deleted by an ``administrator''. 

Over the last few years, the idea of using DLT/blockchains for BIM has been discussed in industry (e.g. in \cite{autodesk-on-blockchain-in-bim,Arup2019}) and in academia. 
Concurrently, the blockchain technology has matured to remove initial shortcomings and limitations: 
for example, enterprise-grade products exhibit a fair amount of performance scalability (see e.g. \cite{gorenflo2019fastfabric}), 
native cloud deployments are available from major cloud vendors (see e.g. \cite{kernahanblockchain}) and
data growth can be contained through checkpointing (see e.g. \cite{fabric-snapshots}) and consensus-controlled deletion (see e.g. \cite{florian-deletion,kuperberg2020deletion}). 

In this technical report, we survey the progress that has been made at the intersection of DLT and BIM, both in industry and in academia. 
Our analysis shows that despite a substantial body of research, commercial adoption is still in very early stages, and there is a lot of unrealized potential despite clear advantages. 

\section{State of the Art: a Survey of Recent Blockchain/BIM Activities}
\label{Related}
\subsection{Products and Commercial Offerings}
\label{products}
As of January 2021, 
almost all publications describe concepts and challenges, but do not create architectures, designs or product-level implementations. 
When it comes to marketed BIM implementation that incorporates DLT or blockchain aspects, we are only aware of BIMCHAIN \cite{Bimchain} but not of any other 
product. 
Likewise, we haven't come across any off-the-shelf blockchain technology or product that incorporates specific BIM aspects such as a Common Data Environment (CDE) or a planning workflow for Model Submission. 

BIMCHAIN is based on Ethereum but otherwise very few technical details are described online (for an independent look at BIMCHAIN, see \cite{pradeep2020}, which is discussed below). 
The FAQ mentions a web interface but does not provide a detailed explanation of the integration with existing tools or file formats. 
Since announcing a beta in 2018, there have been no updates, testimonials or references from the vendor (Lutecium SAS, based in Paris/France).

\subsection{Research Projects and Non-Survey Publications}
\label{projects}
In the following, we analyze recent publications (in alphabetical order), from 2019 and later. 

In \cite{akbarieh2020bim}, 
Akbarieh et al. 
assess blockchain technology as a method to securely store building log-book data which contains all information about building's elements and materials following a ``smart material passport'' approach. 
The goal would be to increase the trustworthiness of material quantities and digital construction data throughout the whole building life cycle. 
The use of smart contracts 
is discussed for providing a digital ``terms and conditions'' equivalent 
and for providing a sound data base that links into BIM, GIS and related Common Built Environment applications. 
The paper qualifies itself as a literature review; no specific implementations of a blockchain-based material passport are being mentioned or referenced.


In \cite{dounas2020a}, 
Dounas et al. present a software prototype that integrates BIM (using Revit) with the Ethereum blockchain technology, utilizing smart contracts. 
They utilize Grasshopper, a visual programming language and environment that runs within the Rhinoceros 3D computer-aided design (CAD) application. 
The authors envision DAO-like collaboration and consensus-based decision making involving the BIM design phase participants. 
The prototype employs IPFS as distributed file storage, but the authors do not discuss the issues of privacy that arises when using IPFS. 
The paper displays several screenshots and design overviews, and the source code of the smart contracts and of the developed plugins is provided in a public Github repository.\footnote{The last sentence in the paragraph has been corrected on May 16th, 2021 to reflect that the source code described in \cite{dounas2020a} is publicly available, see reference 29 in the original paper.} 


In \cite{fitriawijaya2019blockchain}, 
Fitriawijaya et al. describe a small Ethereum-based prototype for supply chain management in the context of BIM. 
They stress the importance of reputation building, in addition to the actual tracking of goods. 
However, their description of the prototype remains abstract, and it appears that it has not been tested publicly. 
The authors do not describes the disadvantages of choosing Ethereum as technology, and do not report whether an integration with commercial BIM tools has been attempted. 

In \cite{hijazi2019}, Hijazi et al. discuss the integration of BIM and blockchains. 
A high-level architecture of the integration is suggested, but no implementation is reported and no technology/product choice is specified. 
The paper also mentions that the authors assessed BIMCHAIN, but does not provide any insights or further findings. 



In \cite{hunhevicz2020-and-smart-contracts},
Hunhevicz et al. propose a financial risk-reward system for the Integrated Project Delivery (IPD). 
A reference to Elghaish et al. \cite{elghaish2020integrated} 
shows the proposed use of Hyperledger Fabric for IPD. 
The authors state that there could be Hyperledger-independent solutions that go beyond financial applications, and that
the digitalization of processes and the creation of incentive mechanisms both allow more automation and increased transparency in project delivery.
A blockchain-based integrated project delivery has the potential to provide an incentive to all stakeholders to optimally manage project resources 
\textcolor{black}
{- an intention that} is in line with the overall project success.
However, no details such as an implementation roadmap or 
design-level refinements are proposed. 
No actual trials have been commenced and the paper does not mention 
\textcolor{black}
{the} downsides to the proposed approach. 

In \cite{hunhevicz2020-crypto-economic},
Hunhevicz and Hall
investigate the integration of digital information with management and contracts.
According to them, such an integration can increase efficiency, transparency and accountability in construction processes between all stakeholders.
The proposed approach utilizes a blockchain-based economic incentive system without laying out any details in regards to a concrete way to implement those.
The concept aims at the weak points of the constrution sector (i.e. lack of productivity, transparency and sustainability) and proposes to tackle them with a blockchain-based incentive system.
There is no implementation referenced by the paper 
and no actual trials have been commenced throughout the research.
The paper does not discuss any possible shortcomings of the proposed idea.  which does not imply that there aren't any.

In \cite{hunhevicz2020-do-you-need}, 
Hunhevicz and Hall 
provide an overview 
of proposed use cases for blockchain technology in the AEC sector, and combine this overview with a decision tree 
to answer the question whether a distributed ledger application is required or not.
The authors provide a matrix that shows which papers look into which use cases.
While many papers and consultancy report exist that talk about \textit{single} use cases, the authors managed to provide an overview on the fragmented research landscape in terms of which of the investigated use cases do require a blockchain-based solution. 
\cite{hunhevicz2020-do-you-need} 
does not propose any concrete implementation or system; correspondingly, 
\textcolor{black}
{no actual trials are described}. 
The paper lists limitations of blockchain based use cases for each of the individual sections 
\textcolor{black}
{that the authors} chose to structure the use case landscape with.
A further paper \cite{hunhevicz2019-managing-mistrust} 
(also from Hunhevicz and Hall) is very similar to \cite{hunhevicz2020-do-you-need}.

In \cite{hunhevicz2020-incentivizing},
Hunhevicz et al. 
propose an Ethereum-based prototype that aims at increasing 
data quantity and quality in construction design projects by using smart contracts and incentive mechanisms.
The 
high-quality data at the end of the design phase often shows to be a problem for the stakeholders trusted with the operation of a building. 
The paper shows an Etherum-based prototype that aims at an increased quality in BIM data sets.
Smart contracts ensure that a project step can only be started once the preceding step (e.g. a sign off of a design) has been completed. 
This approach goes hand-in-hand with an incentive system intrinsic to the proposed solution. 
Stakeholders are not only required to complete steps but also incentivised to do so by the issuance of currency, reputation or securities.
The paper shows how incentives and smart-contract based processes can increase the data quantity and quality in construction design projects.
The authors acknowledge the current cost associated with the use of the Ethereum blockchain mainnet and also agree that other distributed ledger solutions can be used as well.


In \cite{lemes2020}, 
Lemeš 
provides an overview on the potential benefits of a blockchain-based approach for collaborative design, such as information security. 
The paper also discusses the time it will likely take the Architecture, Engineering and Construction sector to adopt Blockchain technology.
The author postulates that in order to stay ahead of their respective competition, Engineering firms are required to adapt to new paradigms and to leverage any available technology so the increasing complexity of modern engineering products doesn't pose a threat to their business.
As modern design teams are often geographically separated, the need for reliable and tracable data is increasing. 
According to Lemeš, Blockchain technology provides an approach to achieve this without the need for a centralized authority that is prone to external attacks in this highly trust-dependent environment.
The paper provides an 
\textcolor{black}
{purely} 
theoretical view on the state of Blockchain technology in AEC and does not feature an actual implementation.
The author mentions relevant down-sides to the blockchain-based approach but still manages to show that the benefits outnumber the disadvantages depending on the use case. 
At the same time, it is apparent that the adoption of Blockchain technology is still in early adoption stage.

In \cite{li2020-framework}, 
Li et al. investigate whether and how pairing Blockchain Technology with BIM, IoT and decentralised autonomous organisations (DAOs) can provide better traceability and better record keeping of products, services and activities in the construction sector, as recommended 
by the UK Government.
The study specifically proposes the integration of BIM, IoT, DLT and smart contracts to achieve 
semi-automated maintenance and repairs of built assets.
By 
augmenting a maintenance step with a human confirmation that the work has been done satisfactorily, the DAO would release a payment and update the data in the CAFM (Computer Aided Facilities Management) system. 
Other information such as certifications or schedules would 
\textcolor{black}
{be added} 
to the information model.
The proposed system 
shall have a positive impact on the reduction of human errors and of administrative overhead by facilities managers, as well as on the productivity and on the creation of new business models.
Additionally, the scalability of DLT and the interoperability of the various systems (DAO, CAFM, blockchain) are mentioned as limitations for the proposed approach.
Unfortunately, the paper does not include an implementation and thus cannot 
\textcolor{black}
{report on the} 
``lessons learned''. 

In \cite{li2019-proposed-approach}, 
Li et al. propose an approach for the integration of DLT, BIM, IoT and Smart Contracts and evaluate a proof-of-concept of a simulated installation task in a BIM-based project using a smart contract.
The smart contract was coded using PyCharm. 
The simulation translates the BIM Employers Information Requirements (EIRs) into smart contract code. 
IoT-based devices measure the quality of the installation and the results trigger (``authorize'') or prevent a payment in the smart contract. 
The proof-of-concept shows how some of the many gaps the overlap of DLT, BIM and IoT might be closed. 
While the paper shows excerpts of the smart contract code, it is not clear whether the code has actually been deployed to a blockchain capable of smart contract execution, and what challenges this would 
\textcolor{black}
{bring} with it.

In \cite{li2019-blockchain-in-the-built-environment}, 
Li et al. look at the current state of DLT in the Built Environment and the construction sector. 
The chosen methodology is a literature review; the authors compiled a list of challenges and opportunities and analysed the regional spread of the origin of reviewed papers. 
Consequently, no actual trials or implementation steps have been in this paper. 

In \cite{li2019-blockchain-in-the-construction-sector}, 
Li et al. propose a multi-dimensional (i.e. social, technological and political) emergent framework to form the basis of DLT adoption within the construction sector.
While the technical dimension deals with the implementation of the technical architecture of DLT, the social dimension looks at the impact DLT will have on society and the political dimension represents the environment in which DLT will be established. 
The authors expect that the proposed framework will be 
\textcolor{black}
{utilized} for a wide range of investigations; 
\textcolor{black}
{the framework does not include code or design}. 
A fourth dimension covering the process aspects is envisioned but not discussed in more detail. 

In \cite{li2019-informing}, 
Li and Kassem provide an overview on the outcome of seven semi-structured interviews with industry practitioners and academics on the topic of DLT in the construction industry. 
The authors acknowledge that the number of interviews (7) provides only a limited perspective of DLT in the construction sector and they conclude that DLT is not a stand-alone solution for the the challenges that the construction industry faces, but that DLT can be a part of the solution.



In 
\cite{nawari2019technologies}, 
Nawari and Ravindran discuss Blockchain technologies for use in the BIM Workflow Environment. 
However, their work does not include an implementation, though they discuss specific technologies such as Hyperledger Fabric. 
In \cite{nawari2019post-disaster}, 
Nawari and Ravindran put DLT/Blockchains and BIM in the context of post-disaster recovery and construction; they present a framework and suggest using Hyperledger Fabric, though no actual implementation is reported. 
In \cite{nawari2019potentials-built-environment}, 
Nawari and Ravindran explore the potentials of blockchains/DLT in BIM and the Built Environment context.  
In \cite{nawari2019review-potential-applications}, 
the same authors consider this topic based on literature research.
At the end, none of the four publications 
\cite{nawari2019technologies,nawari2019post-disaster, nawari2019potentials-built-environment, nawari2019review-potential-applications}
presents an implementation or a detailed implementation design. 

In \cite{peherstorfer2019bim}, 
Peherstorfer combines BIM models storage in a Git repository (which itself uses cryptographic hashes to ensure tamper-proofness of commits), with a release management workflow that is realized as an Ethereum smart contract.
Peherstorfer's resulting \texttt{ChangeManager} is a workflow that claims transparency, traceability/accountability and immutability of persisted data; he has also created a prototype (with a C\# WPF frontend) and compares it to blockchain-less solutions concerning costs and security.
The downside of using Git is that it requires non-IT personnel to understand the Git concepts, which can be complex even for experienced programmers. 
Also, the author discusses the high costs and volatility of Ethereum gas if the approach would be deployed on the public mainnet without storing large model files on cheap Git hosting; tool maturity is another aspect where Peherstorfer sees a need for improvement.

In \cite{pradeep2019}, 
Pradeep et al. perform a literature review (looking at publications from 2018 and earlier), and summarize both the chances and the limitations of Blockchain technology in construction. 
Among other topics, their survey picks up the relationship between the CDE (Common Data Environment) repository and the role of the DLT as a repository. 
However, DLT advances in term of performance, scalability, efficiency, privacy are not discussed in \cite{pradeep2019}, and the authors do not cover any DLT product beyond Ethereum and IOTA. 

In \cite{pradeep2020}, 
Pradeep et al. analyze the beta version of the BIMCHAIN product (cf. Sec.~\ref{products}), based on a list of eight essential functional requirements that the authors set up in the paper: 
	(1) enable a traceable (time-stamped and multi-signature) exchange of data; 
	(2) enable the recording of interactions protected against unauthorized access by \textit{third-party} access (incl. tampering); 
(3) ensure that the interactions records are auditable and transparent; 
(4) protect the records from unauthorized modification even when the modification attempts originate from ``insiders'', i.e. not from external ``third parties''; 
(5) enable the allocation of Intellectual Property rights, and additionally assigning and controlled transfer of ownership to material and immaterial assets; 
(6) ensure that off-chain storage is secure not just in terms of proof-of-integrity, but also protects against data loss which is more likely if the off-chain data is neither replicated nor distributed nor decentralized; 
(7) evaluation of the data and the processing of events shall happen through trustworthy smart contracts which codify criteria, rules and postconditions; and
(8) data from IoT devices and sensors 
\textcolor{black}
{shall be} recorded ``autonomously'', i.e. the BIM model self-updates based on the data received. 
The analysis is performed through desk research and not in a real construction/design project. 
Of the above eight requirements, Pradeep at al. say that 
BIMCHAIN as capable of only four to five.  


In \cite{siountri2019towards},
Siountri et al. propose a high-level system architecture encompassing Blockchain, IoT and Cloud computing for solving informational and security challenges within the environment of museums. 
However, the proposed architecture has not been implemented; the paper focuses on the role of blockchain for  access privileges and traceability. 

In \cite{sreckovic2020analysis}, 
Sreckovic et al. 
perform a literature review 
and discuss how BIM workflows could benefit from the decentralized nature of blockchain-based applications (i.e. ``DApps''). 
The authors propose a conceptual model for the implementation of a blockchain-based and DApps-using design process. 
A concrete implementation 
is not being referenced; 
the development of a design-phase framework for model-based communication is mentioned but not specified. 
The authors do reference an ongoing research project called ``BIMd.sign'' (``BIM digitally signed with blockchain in the design phase'') funded by the Austrian Research Promotion Agency (FFG) and to be finalized by July 2022. 

In \cite{tezel2020preparing}, 
Tezel et al. 
set up the requirements for a construction supply structure facilitated by blockchain technology. 
The authors use qualitative methods (i.e. 17 interviews with subject matter experts) and SWOT analysis to explore the level of preparedness of construction supply chains.
With its focus on investigation, the paper does not provide any technical designs and does not report on implementations. 

In \cite{xue2020semantic}, 
Xue and Lu 
present a semantic differential transaction (SDT) approach which aims at reducing the information redundancy that blockchain-enabled BIM setups tend to bring with them. 
Instead of storing an entire BIM model or its unique digital fingerprint on the blockchain, the SDT approach captures local changes to the model and assembles them into a BIM change contract (BCC). 
This allows for an easy synchronization of BIM changes as the entire version history of a BIM project is a chain of timestamped BCCs. 
The authors use IFC (Industry Foundation Classes - a vendor-agnostic standard)
as the data exchange format for BIM projects and by doing so, show the 
\textcolor{black}
{compatibility} of their SDT approach 
\textcolor{black}
{with} IFC. 
The authors tested their model on two BIM projects using Autodesk Revit 2018, the IFC file format and a ``minimal'' blockchain (i.e. \cite{xue2020semantic} does not use an established enterprise-grade product such as Hyperledger Fabric, Corda or Quorum). 
In essence, the information stored on a blockchain is significantly reduced by capturing changes instead of entire models.
From these delta changes, each participant reconstructs locally (i.e. in their specific BIM software installation) the full model. 
\cite{xue2020semantic} does not report whether the approach was tested in a real-world setting outside of academic research. 

In \cite{wan2020blockchain}, 
Wan et al. 
propose future work for researchers 
after concluding that the reviewed papers contain 
\textcolor{black}
{(at the time when \cite{wan2020blockchain} was written)} no concrete implementations or results thereof.

In \cite{ye2020-integrating},  
Ye et al. propose a framework for an automated management of contracts, invoices and billing. 
Their work is part of a multi-party research project called BIMcontracts \cite{BIMcontracts}. 
In \cite{ye2020-integrating}, the authors analyze the payment logic and the BIM models, but the paper does not showcase a blockchain implementation or a complete implementation of a smart contract. 
Implementation details are extensively covered in the underlying master thesis \cite{ye2019-combining} of Ye, which provides an in-depth explanation of the Fabric-based realization. 
The thesis includes sequence diagrams, setup instractions, screenshots and an extensive analysis of the performed work. 

In \cite{zheng2019bcbim}, 
Zheng et al. argue that BIM has to be combined with other emerging technologies such as Blockchain, mobile cloud, Big Data and IoT. 
Furthermore, a BIM-as-a-service model is described utilizing the aforementioned technologies. 
The authors see the role of blockchain mostly as a technology supporting audit and provenance aspects for BIM. 
A new BIM model called ``bcBIM'' is proposed by the paper with no mention of a concrete implementation of that model. 

\subsection{Survey Publications}
\label{surveys}
In \cite{yang2020public}, 
Yang et al. present a very 
\textcolor{black}
{thorough}, systematic review of blockchain technology not just in BIM, but in construction business process in general. 
In particular, they survey and assess the ``level of adoption'' and classify the studied works as 
``Inception'' (23 out of 27 works) and ``Proof-of-concept'' (the remaining 4 works), as of May 2020 when the survey was compiled. 
Among these 27 publications, 13 deal with BIM; 3 out of 27 deal with a combination of BIM, DLT and IoT/smart sensors and others deal with a combination of IoT and/or RFID with DLT but without BIM. 
It should be noted that \cite{yang2020public} does not cover the publications/projects that we have covered 
\textcolor{black}
{above}
(\textcolor{black}
{specifically} \cite{akbarieh2020bim,
	BIMcontracts,
	dounas2020a,  
	fitriawijaya2019blockchain,
	hijazi2019,
	hunhevicz2020-and-smart-contracts,
	hunhevicz2020-crypto-economic,
	hunhevicz2020-do-you-need,
	hunhevicz2020-implications,
	hunhevicz2020-incentivizing,
	kasten2020engineering, 
	lemes2020,
	li2020-framework,
	li2019-blockchain-in-the-built-environment,
	li2019-blockchain-in-the-construction-sector,
	li2019-proposed-approach,
	li2019-informing, 
	nawari2019technologies, 
	nawari2019post-disaster, 
	peherstorfer2019bim, 
	pradeep2019, 
	pradeep2020, 
	siountri2019towards,
	sreckovic2020analysis,
	tezel2020preparing,
	wan2020blockchain,
	xue2020semantic,
	zheng2019bcbim,
	ye2019-combining, 
	ye2020-integrating}), with the exception of 
\cite{hunhevicz2019-managing-mistrust},
\cite{nawari2019potentials-built-environment} 
and 
\cite{nawari2019review-potential-applications} 
which \cite{yang2020public} mentions 
\textcolor{black}
{only} briefly. 
The following relevant
papers from 2019 and later are covered by Yang as well: 
\begin{itemize}
	\item In \cite{digiuda2020construction}, 
	Di Giuda et al. 
	highlight the benefits of the use of Blockchain Technology in BIM and show smart contracts as a secure and effective means of stakeholder interaction throughout the design, bid, construction and operation phases.
	
	\item In \cite{elghaish2020integrated}, 
	Elghaish et al. 
	present a framework for the integrated project delivery in the AEC sector, 
	which includes an integration into existing BIM processes. 
	The authors managed to build a web-based proof-of-concept providing a platform for financial transactions among design and construction project participants and by doing so go beyond the conceptual level through exploring a case-project proving the concept.
	
	\item 
	In \cite{liu2019building}, 
	Liu et al. look at Blockchain Technology and BIM from a sustainable design perspective and proposs the use of user-level driven smart contracts to enhance BIM systems to achieve better results in the context of sustainable building process. 
	A Building Information Modeling and blockchain (BIM + BC) Sustainable Design Framework is outlined in the paper. 
	No mention of an existing implementation of said framework 
	\textcolor{black}
	{is included in} the paper. 
	
	\item
	In \cite{lokshina2019application}, 
	Lokshina et al. look at Blockchain Technology as a complementary development to BIM and IoT in the context of smart buildings and their operation. 
	\textcolor{black}
	{The authors see benefits 
	in the areas of construction efficiency, 
	safety of humans, 
	and the security of data/information.} 
	No real-world implementation to such a complementary system is mentioned in the paper.
	
	\item In \cite{shojaei2020}, 
	Shojaei et al. present a small use case that 
\textcolor{black}
{they plan to implement 
	using the  Hyperledger Fabric permissioned Blockchain; no current implementation is reported, though.} 
	The authors do not discuss existing industry offerings, such as BIMCHAIN.
	
\end{itemize}

In \cite{kasten2020engineering}, 
Kasten 
provides a comprehensive literature review of blockchain in engineering and manufacturing research papers. 
Most of the reviewed papers see the implementation of their findings as an important next step but cannot look back at any implementation themselves. 
Of the 40+ publications covered in this technical report, Kasten only covers six:  
\cite{
	nawari2019review-potential-applications,
	nawari2019post-disaster,
	nawari2019technologies,
	li2019-blockchain-in-the-built-environment,
	liu2019building,
	zheng2019bcbim
}. 

An overview of the potentials and roles of Blockchain for the built environment from the perspective of an engineering bureau
is presented in \cite{Arup2019,Arup2019supplement}. 

\section{Conclusion}
With several proof-of-concept and prototypic implementations appearing in 2020, the application of DLT/blockchain to BIM topics sees a further increase in interest. 
Originally 
\textcolor{black}
{driven} by academic researchers, we have found that consortial projects increasingly include industrial partners with real-world BIM deployments. 
Interestingly, 
\textcolor{black}
{very few} of the surveyed publications refer to enterprise-grade R3 Corda or Quorum ledgers, and other technologies such as Hashgraph Hedera or Algorand are not mentioned at all. 
When it comes to specific products and implementations, the authors look at Hyperledger Fabric and (generic) Ethereum. 

The activities described in the surveyed publications are mostly kickstarted by groups in the AEC domain (architecture/engineering/construction), and the choice of the 
specific blockchain technology is not a top priority - the efforts overwhelmingly focus on aligning BIM processes and data with the architectural aspects of a DLT-based collaboration setting. 

\textcolor{black}
{
In four 
papers (\cite{dounas2020a,ye2019-combining,peherstorfer2019bim,xue2020semantic}), 
implemented prototypes are described, and thus valuable insights into the practicality of a blockchain-enabled BIM approach are provided. 
\cite{ye2019-combining} uses Hyperledger Fabric and describes the implementation in great detail. 
\cite{dounas2020a} uses Ethereum and Revit (connected using the Rhino framework); 
\cite{peherstorfer2019bim} uses Git for BIM model storage as an intermediate replacement for a blockchain, and 
\cite{xue2020semantic} uses a ``minimal blockchain'' (simulator) for a proof-of-concept. 
}


\bibliographystyle{IEEEtran}
\bibliography{Kuperberg2021c-BIM-Blockchain-Arxiv}

\end{document}